\documentclass[twocolumn,tighten]{aastex62}
\usepackage{natbib,color,amsmath}
\usepackage[normalem]{ulem}
\usepackage{epstopdf}
\usepackage{hyperref}
\usepackage{graphicx}
\definecolor{red}{rgb}{1.0,0.0,0.0}

\newcommand{\Mj}[1]{$M_\mathrm{Jup}$}

\usepackage[utf8]{inputenc}

\NewPageAfterKeywords

\begin{document}
\title{The Possible Astrometric Signature of a Planetary-mass Companion to the Nearby Young Star TW Piscis Austrini (Fomalhaut B): Constraints from Astrometry, Radial Velocities, and Direct Imaging}

\author[0000-0002-4918-0247]{Robert J. De Rosa}
\affiliation{Kavli Institute for Particle Astrophysics and Cosmology, Stanford University, Stanford, CA 94305, USA}

\author[0000-0002-0792-3719]{Thomas M. Esposito}
\affiliation{Department of Astronomy, University of California, Berkeley, CA 94720, USA}

\author[0000-0001-8058-7443]{Lea A. Hirsch}
\affiliation{Kavli Institute for Particle Astrophysics and Cosmology, Stanford University, Stanford, CA 94305, USA}

\author[0000-0001-6975-9056]{Eric L. Nielsen}
\affiliation{Kavli Institute for Particle Astrophysics and Cosmology, Stanford University, Stanford, CA 94305, USA}

\author[0000-0002-5251-2943]{Mark S. Marley}
\affiliation{NASA Ames Research Center, MS 245-3, Mountain View, CA 94035, USA}

\author[0000-0002-6221-5360]{Paul Kalas}
\affiliation{Department of Astronomy, University of California, Berkeley, CA 94720, USA}
\affiliation{SETI Institute, Carl Sagan Center, 189 Bernardo Ave.,  Mountain View CA 94043, USA}

\author[0000-0003-0774-6502]{Jason J. Wang}
\altaffiliation{51 Pegasi b Fellow}
\affiliation{Department of Astronomy, California Institute of Technology, Pasadena, CA 91125, USA}

\author[0000-0003-1212-7538]{Bruce Macintosh}
\affiliation{Kavli Institute for Particle Astrophysics and Cosmology, Stanford University, Stanford, CA 94305, USA}


\begin{abstract}
We present constraints on the presence of substellar companions to the nearby ($d\sim7.6$\,pc), young ($440\pm40$\,Myr) K4Ve star TW Piscis Austrini, the wide ($\sim$0.3\,pc) companion to the A4V star Fomalhaut. We combined absolute astrometry from {\it Hipparcos} and {\it Gaia} with literature radial velocity measurements and dedicated high-contrast imaging observations obtained with Keck/NIRC2 to achieve sensitivity to brown dwarf and planetary-mass companions ($\gtrsim2$\,M$_{\rm Jup}$) over many decades of orbital period ($\lesssim 10^3$\,yr). The significant astrometric acceleration measured between the {\it Hipparcos} and {\it Gaia} catalogues, reported previously in the literature, cannot be explained by the orbital motion of TW PsA around the barycenter of the Fomalhaut triple system. Instead, we find that it is consistent with the reflex motion induced by an orbiting  substellar companion. The combination of astrometry, radial velocities, and a deep $L^{\prime}$ imaging dataset leads to a constraint on the companion mass of $1.2_{-0.6}^{+0.7}$\,M$_{\rm Jup}$. However, the period of the companion is poorly constrained, with a highly multi-modal period posterior distribution due to aliasing with the 24.25 year baseline between {\it Hipparcos} and {\it Gaia}. If confirmed through continued astrometric or spectroscopic monitoring, or via direct detection, the companion to TW PsA would represent a choice target for detailed atmospheric characterization with high-contrast instruments on the upcoming {\it James Webb Space Telescope} and {\it Wide Field Infrared Survey Telescope}.
\end{abstract}

\section{Introduction}
The field of astrometric detection of exoplanets will soon be transformed with the discovery of thousands of planetary-mass companions via the astrometric reflex motion induced on their host star measured by {\it Gaia} \citep{Perryman:2014jr}. While these discoveries will span a wide range of orbital periods, a small subset of them will be at separations wide enough, and around stars young enough, that they will be amenable to direct detection and characterization with ground- and space-based instrumentation. Previous ground-based surveys searching for the astrometric signal induced by an orbiting planet have typically targeted fainter stars due to the requirement to have similar-magnitude reference stars (e.g., \citealp{Sahlmann:2014hu}), whilst space-based observations have focused on characterizing known systems (e.g., \citealp{Benedict:2017gc}). These observations are typically expensive, and dedicated astrometric missions such as {\it Gaia} are required to have sufficient sensitivity around a large enough number of stars for a blind search to be worthwhile. While we await the release of the final {\it Gaia} astrometric catalog, the intermediate data releases (e.g. \citealp{GaiaCollaboration:2018io}) can be combined with previous measurements from {\it Hipparcos} to search for planetary-mass companions to nearby stars (e.g. \citealp{Brandt:2018dj,Kervella:2019bw}).

\object{TW Piscis Austrini} (TW PsA) is a K4Ve star \citep{Keenan:1989jj} at a distance of 7.6\,pc \citep{GaiaCollaboration:2018io}. It shares a similar three-dimensional velocity to the naked-eye star \object{Fomalhaut} \citep{Luyten:1938fs} and the faint M-dwarf \object{LP 876-10} \citep{Mamajek:2013tr}, forming a wide triple system that extends several degrees across the sky. The Fomalhaut system is young, with an estimated age of $440\pm40$\,Myr \citep{Mamajek:2012ga}. This system has been well-studied given its proximity. The primary hosts an extended debris disk \citep{Holland:1998et}, within which a candidate exoplanet with a circumplanetary dust ring has been detected via direct imaging at optical wavelengths \citep{Kalas:2008cs}. LP 876-10 also hosts a debris disk \citep{Kennedy:2014fp}, unusual given the apparent rarity of debris disks around M-dwarfs. Thermal-infrared measurements of TW PsA between 24 and 100\,$\mu$m do not any show evidence of emission in excess of the predicted photospheric flux \citep{Carpenter:2008jb,Kennedy:2014fp}. \citet{Shannon:2014ha} propose that all three stars originate from the same birth cluster, with Fomalhaut and LP 876-10 forming as a binary and TW PsA captured from the cluster into a weakly bound orbit.  While Fomalhaut shows evidence of a planetary system \citep{Kalas:2008cs}, searches for giant planets around TW PsA via direct imaging \citep{Heinze:2010dm,Durkan:2016ib,Nielsen:2019td} and radial velocity \citep{Butler:2017km} have thus far yielded nothing.

In this paper we present constraints on the presence of wide-orbit giant planets around TW PsA with a joint analysis of absolute astrometry from {\it Hipparcos} and {\it Gaia}, a deep high-contrast imaging dataset, and a long-term radial velocity record. We discuss the measured astrometric acceleration, previously noted by \citep{Kervella:2019bw}, in Section~\ref{sec:accl}, where we demonstrate it cannot be explained by the orbit of TW PsA around Fomalhaut, and present limits on the period and mass of an orbiting companion consistent with the measured acceleration. The long-term radial velocity record is presented in Section~\ref{sec:rv}, and the deep coronagraphic imaging dataset in Section~\ref{sec:imaging}. Finally, we present joint constraints on the properties of a companion consistent with the astrometric signal in Section~\ref{sec:joint}.

\section{Astrometry}
\label{sec:accl}
Absolute astrometry for TW PsA was included in both the {\it Hipparcos} (HIP 113283; \citealp{ESA:1997ws,vanLeeuwen:2007dc}) and the recent {\it Gaia} (Gaia DR2 6604147121141267712; \citealp{GaiaCollaboration:2018io}) catalogues. In the {\it Hipparcos} catalogue, the star appears typical compared to other stars within a factor of two in $H_{\rm p}$-band flux. The catalogue goodness of fit metric is $1.33$ compared to $1.30_{-1.39}^{+2.19}$ for the similar brightness stars. This metric is equivalent to $\chi^2=99.5$ for TW PsA, ($\chi^2=131_{-47}^{+71}$ for the comparison sample), corresponding to a $\chi^2_{\nu}=1.2$ given the 87 one-dimensional astrometric measurements of the star and the five model parameters. The quality of fit was good enough that no attempt was made to fit either additional acceleration terms or a stochastic solution \citep{vanLeeuwen:2007du}.

We performed a similar check of the quality of the astrometry in the {\it Gaia} Data Release 2 catalogue. We selected stars within a factor of two in $G$-band flux, that had good quality parallax and photometric measurements as in \citet{Lindegren:2018gy}. The goodness of fit metric for TW PsA was unusually low at 4.7 compared with $19.1_{-9.7}^{+35.1}$ for stars in the comparison sample, indicating a relatively good quality of fit of the five parameter model to the astrometric measurements ($\chi^2=136$ for TW PsA and $\chi^2=960_{-530}^{+3560}$ for the comparison sample). The reduced $\chi^2$ is somewhat high, $\chi^2_{\nu}=2.0$ ($\chi^2_{\nu}=4.3_{-1.9}^{+15.7}$ for the comparison sample), suggesting that the uncertainties on the individual astrometric measurements are slightly underestimated. No significant astrometric excess noise was reported in the catalogue.

\subsection{Astrometric acceleration}
The acceleration of TW PsA between the {\it Hipparcos} and {\it Gaia} missions was estimated by computing three proper motion differentials; (1) $\mu_{\rm G} - \mu_{\rm H}$ the difference between the two catalogue proper motions, (2) $\mu_{\rm H} - \mu_{\rm HG}$ the difference between the proper motion computed from the absolute position of the star in the two catalogues and the {\it Hipparcos} proper motion, and (3) $\mu_{\rm G} - \mu_{\rm HG}$ the same but with respect to the {\it Gaia} proper motion. The proximity of TW PsA (7.6\,pc) causes a non-negligible change in its apparent proper motion across the sky simply due to perspective effects. Without correcting for these, the estimated position of the star at the {\it Gaia} epoch using {\it Hipparcos} astrometry will be systematically offset by 0.2\,mas, biasing the derived $\mu_{\rm HG}$ proper motion by $\sim$10\,$\mu$as\,yr$^{-1}$. The magnitude of this effect is comparable to the size of the uncertainty on $\mu_{\rm HG}$ given the precision of the absolute astrometric measurement of the star within both catalogues.

We account for this effect using the prescription laid out in \citet{Butkevich:2014jt}, propagating the {\it Hipparcos} astrometry from the {\it Hipparcos} to the {\it Gaia} epoch (and vice versa) before computing the difference. Rather than simply dividing the offset between the {\it Hipparcos} and {\it Gaia} coordinates by the 24.25\,yr baseline, $\mu_{\rm HG}$ was instead calculated numerically using a Monte Carlo algorithm given the measurement and uncertainty on the absolute positions of the star at the two epochs and the propagation formalism described in \citet{Butkevich:2014jt}. We assumed that the parallax and radial velocity of the star did not change significantly over the 24.25-year baseline. 

\begin{figure*}
\includegraphics[width=1.0\textwidth]{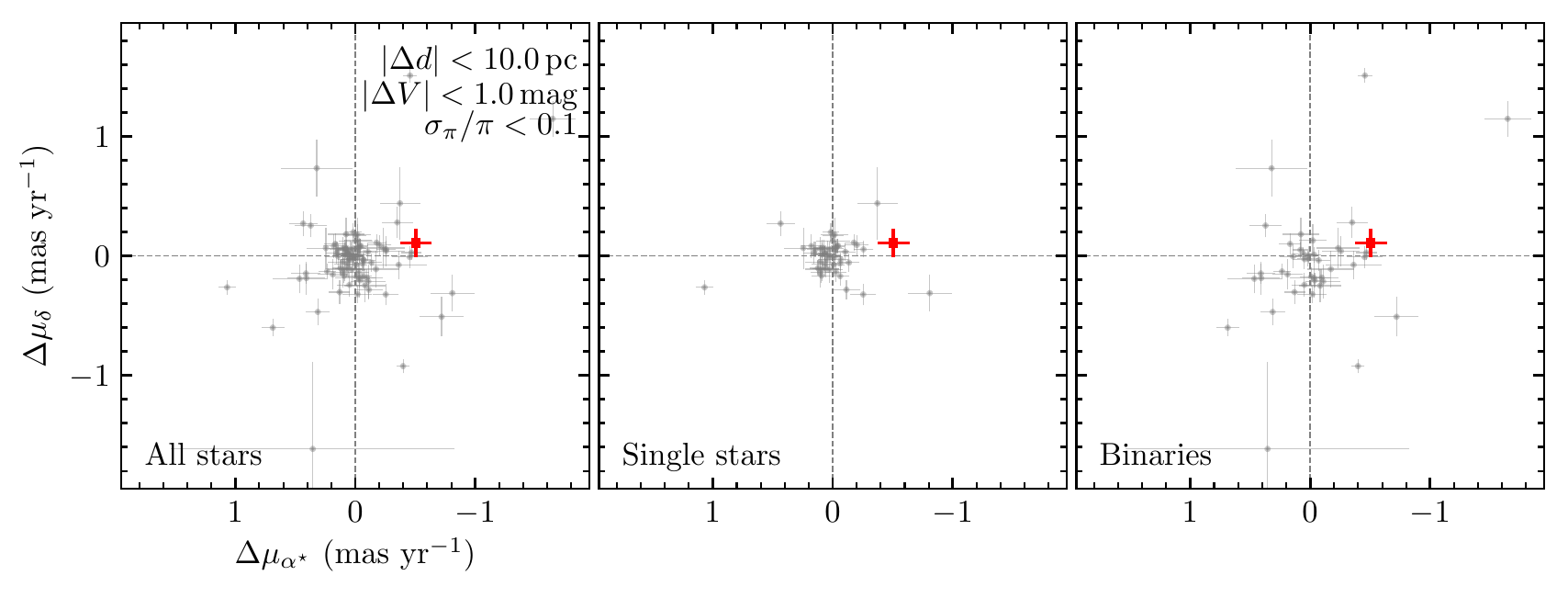}
\caption{The acceleration measured using the {\it Gaia} proper motion $\mu_{\rm G}$ and that derived from the absolute position in the {\it Hipparcos} and {\it Gaia} catalogues $\mu_{\rm HG}$ for a sample of stars sharing similar properties to TW PsA (red symbol). The three panels show all stars (left), those that are thought to be single (middle), and those with evidence of binarity (right). The binary sample contains stars with proper motion differentials of up to 140\,mas\,yr$^{-1}$ (17 of the sample are outside of the plotting window), whereas no star in the single sample has an amplitude greater than 1.1\,mas\,yr$^{-1}$. \label{fig:pm_comparison}}
\end{figure*}
We performed a similar analysis on 106 nearby ($<17$\,pc) stars that have good quality {\it Hipparcos} parallax measurements ($\sigma_{\pi}/\pi < 0.1$) and that share a similar apparent magnitude to TW PsA ($|\Delta V|< 1$\,mag). We plot the $\mu_{\rm G} - \mu_{\rm HG}$ proper motion differential---the most precisely determined---for this sample in Figure~\ref{fig:pm_comparison}. We created two sub-samples based on the binarity of each of the 106 stars; one consisting of the 53 stars that show some evidence of binarity in the literature (either from stellar, substellar, or degenerate companions), and another with those that to the best of our knowledge are single. TW PsA appears to be more discrepant with the majority of the stars in the single sub-sample that have accelerations clustered tightly around zero. The other outliers within the single star sample are likely host to massive companions that have yet to be discovered that are inducing an astrometric acceleration.

\subsection{Orbital motion?}
\label{sec:orbit}
\begin{figure}
\includegraphics[width=1.0\columnwidth]{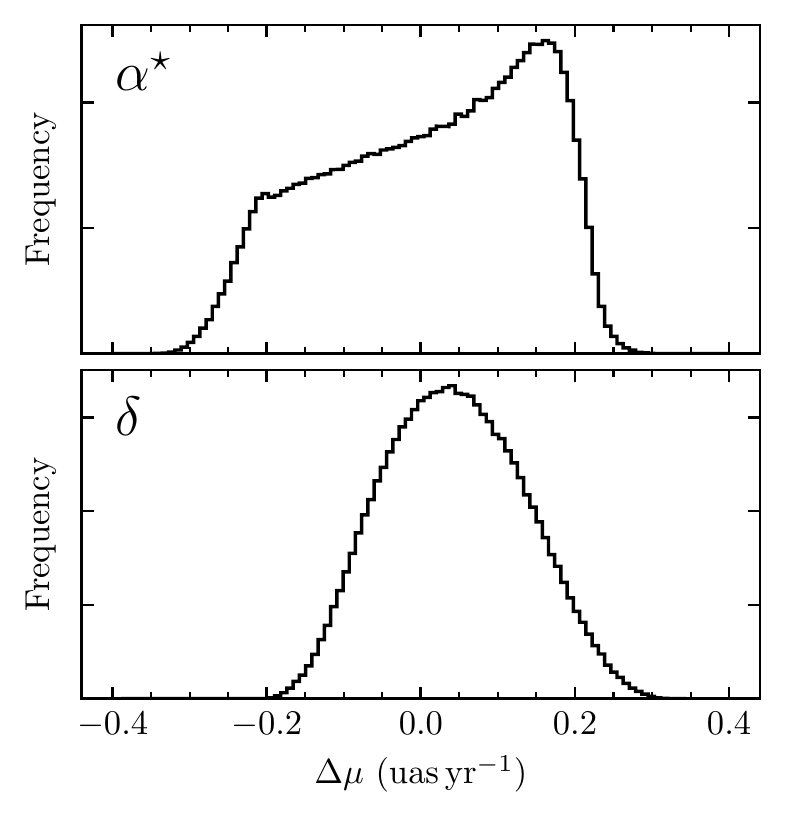}
\caption{Change in the proper motion of TW PsA between the {\it Hipparcos} and {\it Gaia} epochs in the $\alpha^{\star}$ (top) and $\delta$ (bottom) directions induced by orbital motion around Fomalhaut, at a three-dimensional distance of $0.280_{-0.012}^{+0.019}$\,pc. The gravitational influence of the distant companion LP 876-10 is assumed to be negligible.\label{fig:orbit_pm}}
\end{figure}
TW PsA's association with Fomalhaut was first suggested by \citet{Luyten:1938fs}, and was later confirmed with {\it Hipparcos} astrometry by \citet{Shaya:2010dt,Mamajek:2012ga}. The current projected separation of the two stars is almost two degrees on the sky, corresponding to a true separation of $0.280_{-0.012}^{+0.019}$\,pc \citep{Mamajek:2012ga}. The third component of this system is the M4 star LP 876-10, located at a projected separation of approximately six degrees (true separation of $0.77\pm0.01$\,pc) from Fomalhaut \citep{Mamajek:2013tr}. Both of these stars lie within the tidal radius of Fomalhaut, and it is assumed that they form a gravitationally bound triple system. The barycenter of this system lies close to Fomalhaut A (1.92\,M$_{\odot}$), given its mass relative to TW PsA (0.725\,M$_{\odot}$; \citealp{Demory:2009hi}) and LP 876-10 (0.18\,M$_{\odot}$).

The orbital motion of TW PsA around Fomalhaut was noted by \citet{Kervella:2019bw} as being a potential source of the astrometric acceleration measured between the {\it Hipparcos} and {\it Gaia} astrometry. The magnitude of the tangential velocity differential ($\Delta v_{\rm tan} = 18.7\pm6.4$\,m\,s$^{-1}$, \citealp{Kervella:2019bw}) seems inconsistent with the change in orbital velocity expected for an orbital period of close to 8\,Myr \citep{Mamajek:2012ga}. The escape velocity from Fomalhaut is estimated to be 210\,m\,s$^{-1}$ \citep{Mamajek:2012ga}, so it would be surprising for the velocity of TW PsA to change by close to 10\% of the escape velocity in a negligible fraction of the orbital period.

We therefore investigated the plausible ranges of astrometric acceleration over the 24.25-year baseline between the {\it Hipparcos} and {\it Gaia} epochs induced by the orbital motion of TW PsA. For simplicity, we ignored the presence of the low-mass companion LP 876-10 and assumed that Fomalhaut and TW PsA formed a binary with no other massive companions in the system. While this assumption will lead to an imprecise determination of the acceleration induced by the orbit of TW PsA around the barycenter of the triple system, it will be sufficient for an order-of-magnitude estimate of the effect. Given the large angular separation between the two stars, we needed to determine their relative positions in a tangent plane within which the visual orbit could be fit. To achieve this we converted {\it Hipparcos} astrometry and literature radial velocities of the two stars into Cartesian coordinates in the ICRS frame. This coordinate system was then rotated so that the $yz$ plane was tangent to the celestial sphere at the mid-point between the two stars ($\alpha=22^{\rm h}57^{\rm m}1^{\rm s}.33$, $\delta=-30^{\circ}35^{\prime}36.63^{\prime\prime}$), with $+x$ pointing away from Earth.

We used rejection sampling \citep{Blunt:2017et,Blunt:2019vq} to identify visual orbits consistent with the tangent plane separation ($\rho=54070\pm160$\,au) and position angle ($\theta=187.8805\pm0.0002$\,deg) of the pair. The tangent plane position and velocity of the companion from these accepted orbits were first rotated back into the standard ICRS frame and then converted into spherical coordinates to determine the relative parallax and proper motion for each accepted orbit. We then performed an additional rejection step to select only those orbits consistent with the relative parallax ($\Delta \pi=1.61\pm0.78$\,mas) and proper motion ($\Delta \mu_{\alpha^{\star}} = 2.16\pm0.82$\,mas\,yr$^{-1}$, $\Delta \mu_{\delta} = 5.69\pm0.59$\,mas\,yr$^{-1}$) of the two stars from the {\it Hipparcos} catalogue. Here we use the notation $\alpha^{\star} = \alpha\cos\delta$.

From these accepted orbits we calculated the expected change in the proper motion of TW PsA between the {\it Hipparcos} and {\it Gaia} epochs due to orbital motion alone. We find a $3\sigma$ upper limit for the amplitude of the acceleration due to orbital motion of 0.3\,$\mu$as\,yr$^{-1}$ in both the $\alpha^{\star}$ and $\delta$ directions (Figure~\ref{fig:orbit_pm}), corresponding to a tangential velocity change of $\Delta v_{\rm tan} < 0.01$\,m\,s$^{-1}$. While the visual orbit is still rather unconstrained due to the long orbital period ($5.8_{-2.2}^{+3.8}$\,Myr, $4.5_{-1.2}^{+1.8} \times 10^4$\,au), the inclination is marginally constrained $i=65_{-29}^{+30}$\,deg, suggestive of a counter-clockwise orbit for TW PsA around Fomalhaut.

\subsection{Inferred companion properties}
\label{sec:astro_model}
The astrometry of TW PsA measured in the {\it Hipparcos} and {\it Gaia} epochs can be used to infer the range of periods and masses that are consistent with the measured astrometric acceleration. The framework is described in detail in De Rosa et al. 2019, {\it submitted}, but a brief summary will be given here. The astrometric model consists of eleven free parameters. Seven define the orbit of the companion that is perturbing the star; the total semi-major axis $a$, the inclination $i$, eccentricity $e$, argument of periastron $\omega$, longitude of the ascending node $\Omega$, epoch of periastron $\tau$ (in fractions of the orbital period), and the mass of the companion $M_2$. We assume a fixed mass for the primary of 0.725\,M$_{\odot}$. The remaining parameters define the proper motion of the barycenter of the TW PsA system ($\mu_{\alpha^\star}$, $\mu_\delta$) and an offset for the photocenter position at the {\it Hipparcos} epoch to account for the catalogue uncertainties. 

The location and instantaneous proper motion of the photocenter was calculated at both the {\it Hipparcos} and {\it Gaia} epochs by combining the constant motion of the barycenter with the photocenter orbit computed from the orbital elements. The displacement between the barycenter and photocenter was calculated using the mass ratio and flux ratio of the two stars. We used an empirical mass-magnitude relation \citep{Pecaut:2013ej} to estimate the absolute magnitude of TW PsA and the companion, if $M_2$ was above the stellar-substellar limit. The flux of companions with $M_2<0.077$\,$M_{\odot}$ was assumed to be negligible. We corrected for perspective effects using the formalism outlined in \citet{Butkevich:2014jt}, assuming a parallax of 131.438\,mas and a radial velocity of 7.217\,km\,s$^{-1}$ \citep{Soubiran:2013jz}. We assume that the parallax and radial velocity are precisely determined to minimize the number of free parameters within the fit.

\begin{deluxetable*}{ccl}
\tablewidth{0pt}
\tablecaption{Measurements and derived quantities in the astrometric model\label{tbl:fit1_params}}
\tablehead{\colhead{Symbol} & \colhead{Unit} & \colhead{Description}}
\startdata
$\Delta\alpha_0^{\star}$, $\Delta\delta_0$ & mas & Offset between barycenter and {\it Hipparcos} catalogue position (1991.25)\\
$\alpha_1^{\star}$, $\delta_1$ & mas & Predicted position of $G$-band photocenter (2015.5)\\
$\alpha_{\rm G}^{\star}$, $\delta_{\rm G}$ & mas & Position of $G$-band photocenter measured by {\it Gaia} (2015.5)\\
$\mu_{\alpha^{\star}}$, $\mu_{\alpha^{\star}}$ & mas\,yr$^{-1}$ & Proper motion of system barycenter (1991.25)\\
$\mu_{\alpha^{\star}, 0}$, $\mu_{\delta, 0}$ & mas\,yr$^{-1}$ & Orbital motion of photocenter around barycenter (1991.25)\\
$\mu_{\alpha^{\star}, 1}$, $\mu_{\delta, 1}$ & mas\,yr$^{-1}$ & Orbital motion of photocenter around barycenter (2015.5)\\
$\mu_{\alpha^{\star}, {\rm H}}$, $\mu_{\delta, {\rm H}}$ & mas\,yr$^{-1}$ & Proper motion measured by {\it Hipparcos} (1991.25)\\
$\mu_{\alpha^{\star}, {\rm G}}$, $\mu_{\delta, {\rm G}}$ & mas\,yr$^{-1}$ & Proper motion measured by {\it Gaia} (2015.5)\\
\enddata
\end{deluxetable*}

As in De Rosa et al. 2019, {\it submitted}, we used the parallel-tempered affine-invariant Markov chain Monte Carlo ensemble sampler {\tt emcee} \citep{ForemanMackey:2013io} to sample the posterior distributions of the eleven model parameters. We computed the likelihood $\ln {\mathcal L} = -\chi^2/2$ at each step in the chain by comparing the predicted position and proper motion of the photocenter at both the {\it Hipparcos} and {\it Gaia} epochs with the measurements from both catalogues as
\begin{equation}
\label{eqn:chi2_1}
    \chi^2 = R_{\rm H}^\top {\mathbf C}_{\rm H}^{-1}R_{\rm H} + R_{\rm G}^\top {\mathbf C}_{\rm G}^{-1}R_{\rm G}
\end{equation}
with H and G subscripts denoting measurements of TW PsA from the {\it Hipparcos} and {\it Gaia} catalogues, respectively. The residual vectors $R$ were calculated as
\begin{equation}
\begin{split}
    R_{\rm H} = &[\Delta\alpha^{\star}_0, \Delta\delta_0,\\
    &\mu_\alpha^{\star} + \mu_{\alpha^{\star},0} - \mu_{\alpha^{\star},{\rm H}}, \mu_\delta + \mu_{\delta,0} - \mu_{\delta,{\rm H}}]\\
    R_{\rm G} = &[(\alpha_1 - \alpha_{\rm G})\cos\delta_1, \delta_1 - \delta_{\rm G},\\
    &\mu_\alpha^{\star} + \mu_{\alpha^{\star},1} - \mu_{\alpha^{\star},{\rm G}}, \mu_\delta + \mu_{\delta,1} - \mu_{\delta,{\rm G}}].
\end{split}
\end{equation}
The variables are described in Table~\ref{tbl:fit1_params}. The covariance matrices ${\mathbf C}_{\rm H}$ and ${\mathbf C}_{\rm G}$ for the {\it Hipparcos} and {\it Gaia} measurements of TW PsA. ${\mathbf C}_{\rm H}$ was computed from the weight matrix ${\mathbf U}$ obtained from the {\it Hipparcos} catalogue using the procedure described in \citet{Michalik:2014eu}, while ${\mathbf C}_{\rm G}$ was computed directly from the correlation coefficients given in the {\it Gaia} catalogue. The column and row corresponding to the covariance between the parallax and the coordinates and proper motion were removed as the parallax of the star was not a free parameter in this fit.

We used a parallel-tempered scheme with 24 steps on the temperature ladder; the lowest temperatures sample the posterior distribution, whilst the highest sample the priors. The walkers at each step of the temperature ladder periodically exchange positions, allowing for a more efficient sampling of a highly multi-modal likelihood surface \citep{Earl:2005iz}. At each temperature we initialized 2048 chains and advanced them for $10^6$ steps, saving every hundredth step. The first half of each chain was discarded as a ``burn-in''. The auto-correlation length of the decimated chains was close to unity, suggesting that each saved sample was independent of the last. Despite the large number of iterations the chains did not appear fully converged. The median and $1\sigma$ ranges of several of the parameters were still slowly evolving when the chains were terminated. This is likely due to the highly multi-modal likelihood surface, and the difficulty in efficiently moving between the distinct areas of high likelihood. We decided not to advance the chains further; the islands of high likelihood in the mass-period plane have all been explored, it is only the relative likelihood that would be better constrained with fully converged chains.

\begin{figure*}
\includegraphics[width=1.0\textwidth]{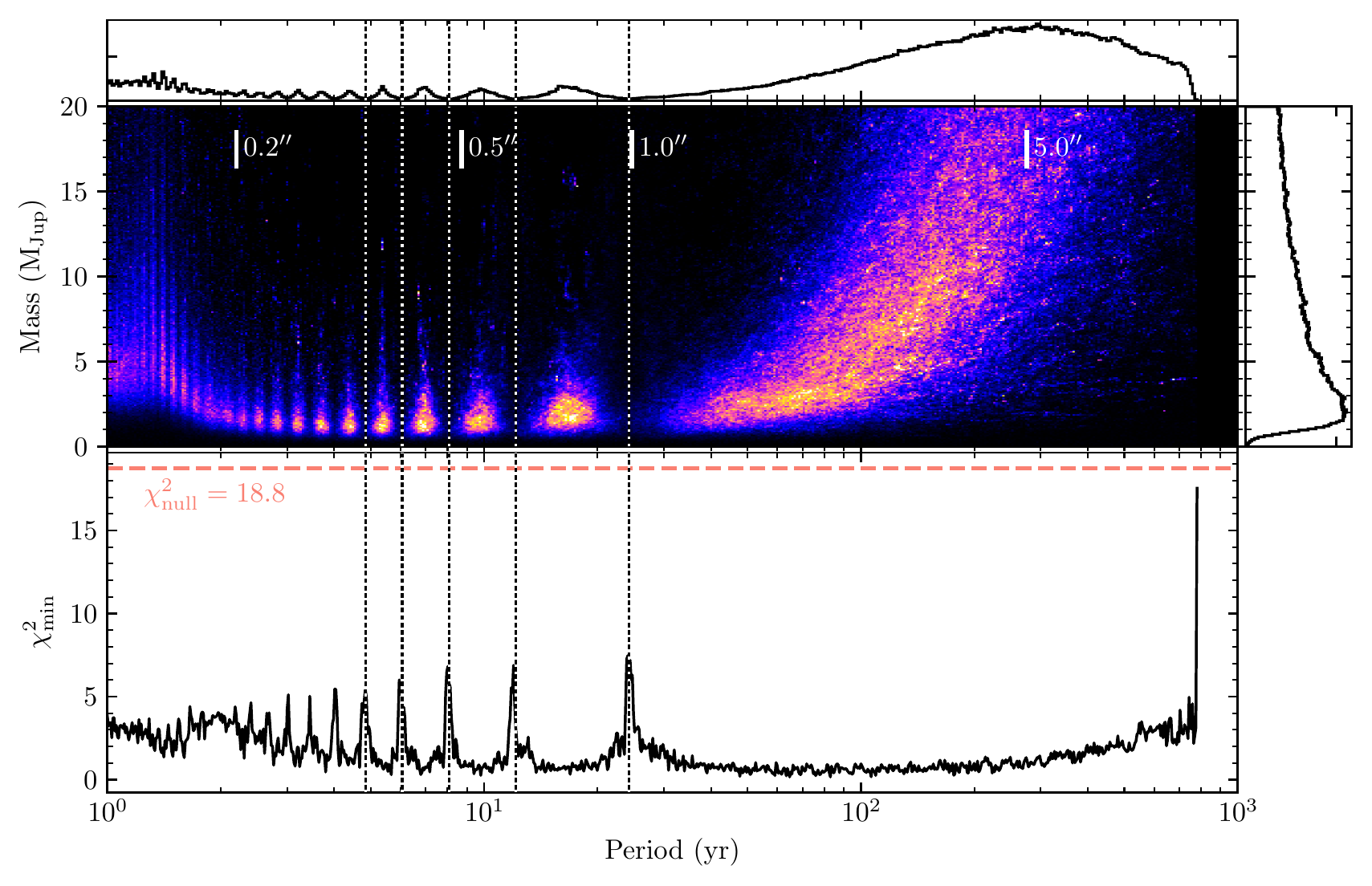}
\caption{Period-mass posterior distribution for companions consistent with the astrometric acceleration of TW PsA measured between {\it Hipparcos} and {\it Gaia} (top panel), with associated marginalized distributions. The minimum $\chi^2$ as a function of period is also plotted (bottom panel), highlighting the aliasing with integer fractions of the 24.25-year baseline between the two missions (dashed vertical lines). \label{fig:instant_prediction}}
\end{figure*}
A two-dimensional histogram of the period and mass of companions consistent with the measured astrometry is shown in Figure~\ref{fig:instant_prediction}, along with their marginalized distributions and a plot of the minimum $\chi^2$ as a function of orbital period. The astrometric signal is consistent with the perturbation of TW PsA by a planetary-mass companion, although the period is almost completely unconstrained. The periodic nature of the high probability regions in the mass-period plane is due to aliasing of the orbit with the 24.25-year baseline between the {\it Hipparcos} and {\it Gaia} missions. At periods greater than 25\,yr the mass and period are strongly correlated; a longer period corresponds to a smaller acceleration over a 25-year baseline given a fixed mass and so a more massive companion is required to explain the observed signal.

The astrometric acceleration is also consistent with near equal mass (and thus equal brightness) companions that, for clarity, are not shown in Figure~\ref{fig:instant_prediction}. The photocenter orbit induced by a near equal brightness companion mimics that of a significantly lower mass companion that contributes negligible flux. A massive stellar companion would induce a significant radial velocity signal, or would be readily identified with high-contrast imaging, depending on its orbital period. It would be surprising if such a well-studied nearby, young star had a stellar companion that had not been discovered despite numerous studies with a variety of instruments. Indeed, we strongly exclude a stellar companion to TW PsA over a wide range of orbital periods based on literature radial velocities (Section~\ref{sec:rv}), and dedicated high-contrast imaging (Section~\ref{sec:imaging}).

\subsection{Null hypothesis}
\label{sec:null1}
The significance of the astrometric acceleration can be tested by repeating the same fit with $M_2$ fixed to zero and comparing the goodness of fit to that of the original model. The only free parameters in this fit are the four astrometric parameters defining the position and motion of the barycenter. We find a $\chi^2_{\rm min}=18.8$ for this simplified model, compared with $\chi^2_{\rm min}=0.05$ found previously. While the change in the $\chi^2_{\rm min}$ suggests that adding a massive companion significantly improves the quality of the fit, the magnitude of the change is not surprising given that the number of free parameters in the model has more than doubled. We therefore used the Bayesian information criterion (BIC) to evaluate the improvement in the quality of the fit. The BIC is defined as $\chi^2 + k\ln n$, where $k$ is the number of parameters in the model, and $n$ is the number of data points. The BIC strongly penalizes the inclusion of additional free parameters in the model to improve the quality of the fit. The $\Delta$BIC between the two models was calculated to be $4.2$ in favour of the eleven-parameter fit, supporting the companion hypothesis but not at a significant level using the categorization from \citep{Kass:1995eh}. The penalty for having more model parameters is particularly apparent from the $\Delta$BIC calculated when considering only near-circular ($e<0.001$) orbits from the fit in Section~\ref{sec:astro_model}. For this subset $e$ and $\omega$ are effectively fixed, reducing the number of free parameters to seven. We find $\chi^2_{\rm min}=0.5$ for this subset, corresponding to $\Delta$BIC=12.0, strong evidence in favour of the companion hypothesis in this restricted scenario.

\section{Radial Velocities}
\label{sec:rv}
TW PsA was noted by \citet{Busko:1978ua} as potentially being a spectroscopic binary, probably due to the range of radial velocities given for the star in the literature at the time (0--15\,km\,s$^{-1}$; \citealp{Popper:1942dn,Joy:1947dh,Evans:1957dv}). No individual investigator noted a variable velocity for the star, although only a handful of measurements were taken by each. Another astrophysical source of radial velocity variations is stellar jitter, a signal induced by the chromospheric activity of young, solar-type stars. TW PsA appears moderately active relative to stars of a similar age, with $\log R^{\prime}_{\rm HK}=-4.5$ to $-4.3$ \citep{Henry:1996dn,Gray:2006ca,Jenkins:2006ja}. A fit to the scatter of radial velocity measurements as a function of stellar activity for a large sample of young stars demonstrates a clear trend of decreasing jitter with decreasing $\log R^{\prime}_{\rm HK}$ \citep{Hillenbrand:2015vp}. Using this empirical relation, the measured activity index for TW PsA suggests a jitter amplitude of $\sigma_v = 25_{-13}^{+25}$\,m\,s$^{-1}$, several orders of magnitude smaller than the discrepancy between the historical radial velocity measurements. More modern radial velocity measurements have provided a more consistent estimate for the velocity of 7.0--7.2\,km\,s$^{-1}$ \citep{Chubak:2011wn, Soubiran:2013jz, GaiaCollaboration:2018io}, evidence against TW PsA being a spectroscopic binary.

\subsection{Literature Keck/HIRES velocities}
\begin{figure}
\includegraphics[width=1.0\columnwidth]{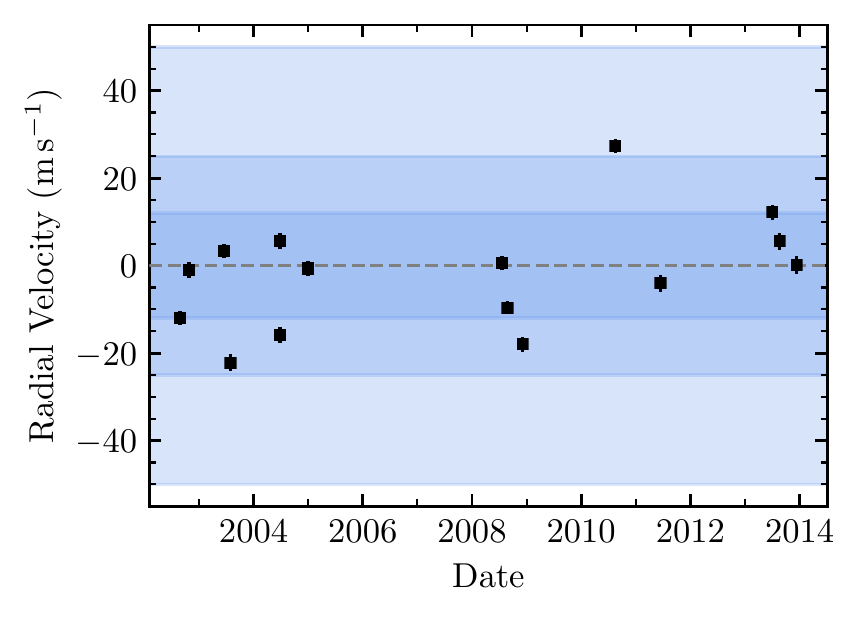}
\caption{Radial velocity measurements of TW PsA obtained with Keck/HIRES over an eleven year baseline \citep{Butler:2017km, TalOr:2019gm}. The weighted average of the velocities was subtracted, centering the measurements around 0\,km\,s$^{-1}$. The amplitude of the stellar jitter estimated from activity indicators is highlighted (dark blue - 16th percentile, blue - median, light blue - 84th percentile). \label{fig:hires}}
\end{figure}
Comparing radial velocities measured using different instruments and different data reduction techniques can be problematic when searching for small-amplitude velocity variations. To mitigate these potential biases, we used spectroscopic measurements of TW PsA taken over a long baseline with the same instrument and reduced and analyzed with the same pipeline. High-resolution optical echelle spectra of TW PsA were taken regularly with the HIRES instrument on the Keck I telescope as part of the Lick-Carnegie Exoplanet Survey \citep{Butler:2017km}. A total of fifteen radial velocity measurements of the star were made between mid-2002 and late-2013 (Figure~\ref{fig:hires}). The velocities were initially published in \citet{Butler:2017km}, but were recently corrected for small systematic errors by \citet{TalOr:2019gm}. While the measurement error on each individual velocity is small, there is a significant scatter about the mean. The amplitude of this scatter is consistent with the amplitude of the stellar jitter expected given the activity indicators measured for TW PsA.

The radial velocities do not exhibit large-amplitude variations induced by an orbiting stellar companion. Face-on orbits are not strictly excluded, however the inclination for a stellar companion would have to be very small. For example, a 0.1\,$M_{\odot}$ companion on a circular orbit with a semi-major axis of 0.1\, au requires an inclination of $i<1^{\circ}$ to induce a velocity semi-amplitude less than $30$\,m\,s$^{-1}$. This limit rises to $i<3^{\circ}$ for a 1\,au orbit, and $i<8^{\circ}$ for a 10\,au orbit, wide enough to be resolved via direct imaging. 

\begin{figure}
\includegraphics[width=1.0\columnwidth]{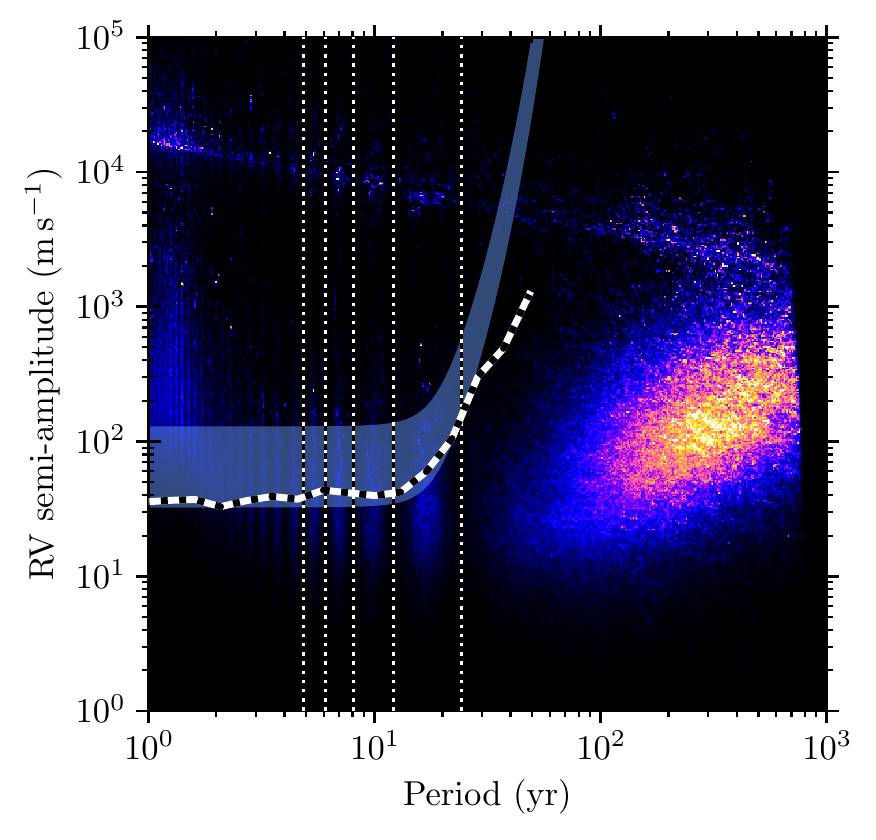}
\caption{Period-RV semi-amplitude posterior distribution for companions consistent with the astrometric acceleration of TW PsA measured between {\it Hipparcos} and {\it Gaia}. A distinct population of stellar companions is seen with a semi-amplitude of $K_1>1$\,km\,s$^{-1}$, as well as a significant population at longer periods and lower semi-amplitudes. The sensitivity of the HIRES RV record calculated using the injection recovery framework is plotted (dashed), as well as an analytical approximation using the framework described in \citet{Howard:2016gc} (blue region denotes possible range of the sensitivity limit). \label{fig:rv_prediction}}
\end{figure}
To assess the sensitivity of these measurements to stellar and substellar companions, we used an injection and recovery scheme written within the framework of the \texttt{radvel} radial velocity analysis code \citep{Fulton:2018gq}. In brief, we simulated 10000 synthetic orbital signatures with orbital periods ranging from 10 to 20000 days and Doppler semi-amplitudes ranging from 10\,m\,s$^{-1}$ to 100\,km\,s$^{-1}$. These broad ranges allowed us to probe companions from the planetary to the stellar mass regime, and out to well beyond the time baseline of the RV data. For each synthetic companion, the RV curve was computed and added to the existing RV data. A least-squares minimization was then performed with period fixed near the injected period and all other parameters allowed to vary. We compared the resulting best-fit solution to a zero-companion solution via the BIC to determine whether the injected companion was recovered. 

This methodology was a simplified version of a full injection and recovery test as described in \citet{Howard:2016gc}, since we checked the solution at only the injected period, rather than computing a full periodogram for each injected companion. We found that the small number of RV observations and sparsity of the cadence prevented traditional recovery from a full periodogram. Because of the complicated window function of the data, the periodogram structure was characterized by strong peaks at many different aliases of the injected period, which overwhelmed the true injected peak, particularly for large-amplitude injections. A real high-mass companion would therefore be evident in the periodogram structure, but would be difficult to characterize from this data alone. Because of this detail, we note that the results of our injection/recovery test likely overestimate our sensitivity to companions. However, the results, plotted in Figure \ref{fig:rv_prediction}, are useful as an estimate for the regions of parameter space in which a companion could still be located. As a conservative estimate, the RV data alone cannot rule out companions with periods $\gtrsim40$\,yr or with RV semi-amplitudes of $<30$\,m\,s$^{-1}$.

\section{Direct Imaging}
\label{sec:imaging}
TW PsA has been observed with several high-contrast imaging systems (e.g., Keck/NIRC2, Gemini South/GPI, VLT/SPHERE) that have sufficient sensitivity to detect stellar and substellar companions over a range of angular separations and masses. For example, \citet{Nielsen:2019td} report a null detection of substellar companions around TW PsA in their Gemini Planet Imager observations, despite having sensitivity to stellar, brown dwarf, and high-mass giant planets between $\sim0\farcs2$ and $1\arcsec$ radius. TW PsA is not an optimal target for instruments such as GPI that operate at near-infrared wavelengths (1--2\,\micron); at $\sim$450\,Myr, giant planets will have cooled sufficiently such that the contrast between star and planet will be more favourable for direct detection at wavelengths between 3\,\micron\ and 5\,\micron.

\subsection{Keck II/NIRC2 observations}
TW PsA was observed on 2012 July 20 with the NIRC2 instrument on the Keck II telescope in conjunction with the facility adaptive optics (AO) system. Observations were conducted with TW PsA serving as its own AO natural guide star and the instrument in vertical angle mode, causing the angle of North to change with the parallactic angle of the target over time. This observing mode enables angular differential imaging (ADI; \citealp{Marois:2006df}), where the point-spread function remains fixed relative to the detector while astrophysical signals rotate as the target transits the observatory. Conditions were photometric with atmospheric seeing ${\sim}0\farcs6$ and precipitable water vapor ${\sim}2$\,mm. The star was placed behind the 100\,mas radius semi-transparent coronagraphic mask (``corona200''), except when offset to obtain a measurement of the sky background. We obtained 49 0.8-s images with 30 coadds, the short exposure time being necessary to both prevent saturation near the edge of the coronagraphic mask and to limit the number of thermal background counts within each image. The field of view rotated by $15.6\deg$ and airmass ranged from 1.60 to 1.69 over the course of the sequence. 48 sky offset frames were obtained at the mid-point and end of the observing sequence. All observations were taken using the $L^{\prime}$ filter (3.4--4.1\,\micron), inscribed circle pupil, and the narrow camera with a $10\arcsec\times10\arcsec$ field of view and 9.952\,mas pixel$^{-1}$ plate scale \citep{Yelda:2010ig}. We did not obtain any images of TW PsA where the star was not saturated. Instead, we observed the $L^{\prime}$ photometric calibrator star G158-27 ($L^{\prime}=6.898\pm0.014$ mag) immediately afterwards, obtaining nine 0.2-s images, with 50 coadds per image, in a three-point dither pattern.

After standard bias subtraction and thermal background subtraction, we masked cosmic-ray hits and other bad pixels. We then aligned individual exposures relative to each other via cross-correlation of their stellar diffraction spikes \citep{Marois:2006df} and the absolute star center was measured via radon transform \citep{Pueyo:2015cx}. Following the procedure described in \citet{Esposito:2014gs}, we applied a LOCI algorithm (``locally optimized combination of images''; \citealp{Lafreniere:2007bg}) to suppress the stellar PSF and quasistatic speckle noise in the data. The reduction presented herein used manually tuned parameters of $N_\delta=0.1$, $W=10$ pixels, $dr=10$ pixels, $g=0.5$, and $N_a=10$ to subtract the region between radii of 11 and 470 pixels, following the conventional parameter definitions in \citet{Lafreniere:2007bg}. Individual PSF-subtracted frames were then derotated to place north up and averaged to produce the final image.

\begin{figure*}
\includegraphics[width=1.0\textwidth]{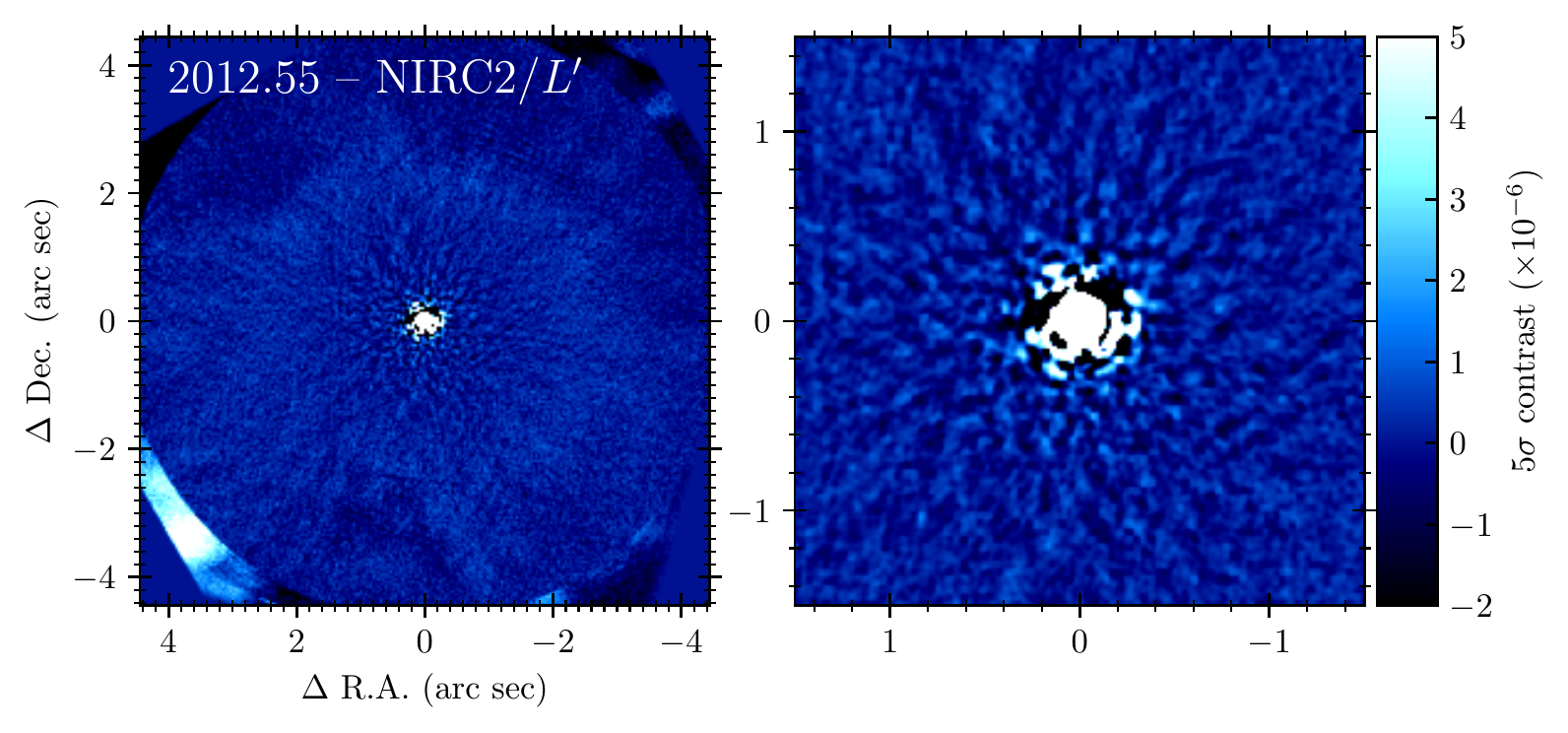}
\caption{Sensitivity to point sources in the vicinity of TW PsA from the NIRC2 $L^{\prime}$ images after PSF subtraction and combination. The two panels show the full field of view (left) and the inner $1\farcs5$ (right). No point source is detected at a significant level within $4\farcs7$. \label{fig:nirc2_im}}
\end{figure*}

\subsection{Companion mass limits}
\begin{figure}
\includegraphics[width=1.0\columnwidth]{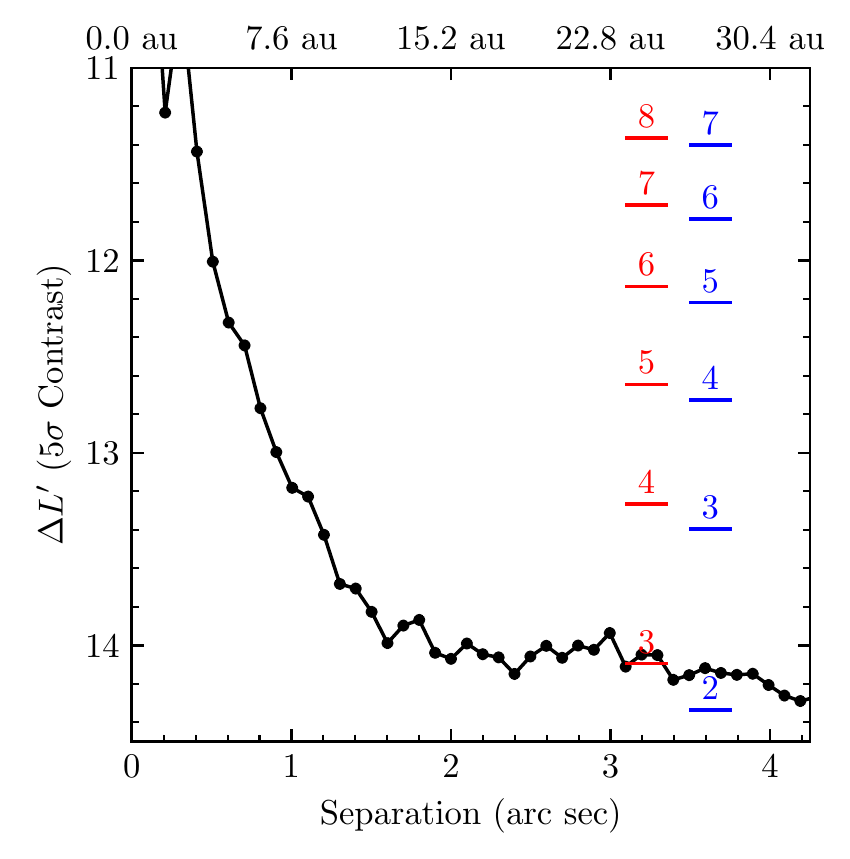}
\caption{Azimuthally averaged $5\sigma$ contrast curve calculated from the final PSF-subtracted image of TW PsA. Companion mass limits (in $M_{\rm Jup}$) are shown, derived from the {\tt Sonora} (red) and {\tt COND03} (blue) evolutionary models assuming an age of 440\,Myr, and an apparent magnitude of $L^{\prime}=3.8$\,mag for TW PsA. Projected separation is shown assuming a distance of 7.6\,pc.\label{fig:nirc2_con}}
\end{figure}

We established our sensitivity to point sources in the NIRC2 data by computing achieved contrast as a function of projected separation. To do so, we first converted the final LOCI image from detector counts to contrast units via the measured flux of the photometric calibrator star (G158-27), assuming the photometric conditions to be constant between the science and calibration observations. We then measured the azimuthally-averaged $5\sigma$-equivalent contrast limits as a function of separation from the star \citep{Mawet:2014ga}. Finally, we corrected the contrast curves for point-source flux attenuation from LOCI PSF subtraction, which we determined by injecting and recovering simulated planets of known brightness across our separation range. The final contrast curve is shown in Figure~\ref{fig:nirc2_con}.

To assess the sensitivity of these observations to planets around TW PsA, we used the {\tt Sonora} grid\footnote{\url{https://zenodo.org/record/1309035}}, a combined evolutionary and atmospheric model grid for cloud-free substellar objects (Marley et al. 2019, {\it in prep.}). A mass-magnitude relationship was created for each filter by performing a linear interpolation of the atmospheric model grid at the predicted temperature and surface gravity of each point within the evolutionary grid. Fluxes at arbitrary planet mass and age could then be calculated by performing a linear interpolation of this new grid. The rapid decline in temperature for substellar objects as a function of their age results in companions $<1$\,M$_{\rm Jup}$ having temperatures lower than the lowest temperature within the grid at 320\,Myr, rising to $<1.5$\,M$_{\rm Jup}$ at 560\,Myr. For the purposes of this study, we assume that planets that do not lie within the boundary of the atmospheric grid have zero flux. For higher-mass brown dwarf and stellar companions we instead use the {\tt COND03} evolutionary models \citep{Baraffe:2003bj}. The mass limits for a given $\Delta L^{\prime}$ from both the {\tt Sonora} and {\tt COND03} evolutionary models are shown in Figure~\ref{fig:nirc2_con}.

\section{Joint Constraints}
\label{sec:joint}
\subsection{Revised astrometric model}
The model described in Section~\ref{sec:astro_model} used to identify TW PsA as potentially hosting a substellar companion makes a simplifying assumption that {\it Hipparcos} and {\it Gaia} both measured the average proper motion of the photocenter over the duration of their respective missions. While this assumption is valid for companions on a long orbital period, where the change in proper motion over a few years is small relative to the astrometric precision, it breaks down for short-period systems that exhibit significant curvature in the photocenter orbit over a few years. For these short-period systems, the average proper motion measured by either {\it Hipparcos} or {\it Gaia} is a strong function of the phasing of the measurements with respect to the photocenter orbit, especially for eccentric systems. The model described in Section~\ref{sec:astro_model} attempts to account for this bias by performing a least-squares fit to the location of the photocenter on the dates that {\it Hipparcos} and {\it Gaia} obtained an astrometric measurement of the star. This approximation assumes that the measurements are equally weighted, and does not account for the one-dimensional nature of the measurement where the location of the star is only constrained along a given direction.

The plausible range of periods for the companion (Figure~\ref{fig:instant_prediction}) motivated us to use the actual astrometric measurements made by the {\it Hipparcos} satellite to constrain the curvature of the photocenter orbit over the {\it Hipparcos} epoch. These observations, the Intermediate Astrometric Data (IAD; \citealp{vanLeeuwen:2007du}), are 87 one-dimensional measurements of the position of TW PsA measured by the {\it Hipparcos} satellite between 1989.96 and 1992.86. The orientation ($\psi$) of these scans relative to the celestial sphere determines how constraining each measurement is in the $\alpha^\star$ and $\delta$ directions, with a typical scan uncertainty of $2.8\pm0.6$\,mas. We reconstructed the original abscissa measurement using the residuals to the five-parameter fit ($\alpha^\star_0$, $\delta_0$, $\pi$, $\mu_{\alpha^\star}$, $\mu_\delta$) given in the IAD, and the catalogue astrometry for TW PsA from the {\it Hipparcos} re-reduction \citep{vanLeeuwen:2007dc}.

The reconstructed abscissa $\Lambda_{\rm HIP}$ can be compared to a model abscissa $\Lambda_{\rm phot}$ that includes the photocenter motion induced by the orbiting companion, calculated as
\begin{equation}
\label{eqn:abs_model}
\begin{split}
    \Lambda_{\rm phot} = &\left[\alpha^\star + t\mu_{\alpha^\star} + x_{\rm orbit}\left(t\right) \right]\cos\psi\\
     &+ \left[\delta + t\mu_\delta + y_{\rm orbit}\left(t\right)\right]\sin\psi\\
     &+ \pi\Pi\left(t\right),
\end{split}
\end{equation}
where $t$ are the scan epochs in years relative to 1991.25, $x_{\rm orbit}$ and $y_{\rm orbit}$ are the position of the photocenter relative to the barycenter, and $\Pi$ is the parallax factor \citep{Sahlmann:2010hh}. Reconstructing the abscissa and performing the fit in two dimensions using the method outlined in Nielsen et al. (2019b, {\it submitted}) produces consistent results. These measurements were incorporated into the model described in Section~\ref{sec:astro_model} by replacing the first term of Equation~\ref{eqn:chi2_1} with $\chi^2_{\rm H}$, calculated using the reconstructed abscissa and Equation~\ref{eqn:abs_model} as
\begin{equation}
    \chi^2_{\rm H} = \left[\left(\Lambda_{\rm HIP} - \Lambda_{\rm phot}\right)/\sigma_{\Lambda_{\rm HIP}}\right]^2.
\end{equation}

The measurements used to construct the {\it Gaia} DR2 catalogue are not due to be released until the conclusion of the nominal mission, so we are not yet able to incorporate the individual {\it Gaia} measurements into our model. The good quality of the astrometric fit relative to stars of a similar magnitude, and the lack of any significant detection of astrometric excess noise, suggests that the assumption of linear motion of the photocenter over the {\it Gaia} epoch is reasonable. A significant detection of curvature in the photocenter motion would lead to a worse $\chi^2$ as all stars within the catalogue were fit using the same five-parameter astrometric model. Once these individual measurements become publicly available, the analysis presented within this work should be repeated. The improvement in the precision of the astrometric measurements relative to {\it Hipparcos} will significantly constrain the mass and orbital properties of any companion to TW PsA.

\subsection{RV and imaging constraints}
The constraints provided by the radial velocity record (Section~\ref{sec:rv}) and the high-contrast imaging dataset (Section~\ref{sec:imaging}) were incorporated into the model by including two additional terms when calculating the goodness of fit; $\chi^2_{\rm RV}$ and $\chi^2_{\rm Img}$. The first was calculated as 
\begin{equation}
    \chi^2_{\rm RV} = \sum_i \left(\frac{v_{i,\rm obs} - v_{i, \rm model}}{\sqrt{\sigma_{v_i}^2 + \sigma_{\rm jitter}^2}}\right)^2 ,
\end{equation}
where $v_{\rm obs}$ are the HIRES radial velocities, $v_{\rm model}$ are the predicted radial velocities (a combination of the reflex motion induced by the companion at each epoch and systemic velocity $\gamma$), $\sigma_v$ are the uncertainties on each HIRES measurement, and $\sigma_{\rm jitter}$ is the amplitude of the radial velocity jitter. Following \citet{Howard:2014bu}, we added the following penalty term to the log likelihood to limit values of $\sigma_{\rm jitter}$ 
\begin{equation}
    -\sum_i\ln\sqrt{2\pi\left(\sigma_{v_i}^2 + \sigma_{\rm jitter}^2\right)}
\end{equation}

The second term added to the goodness of fit, $\chi^2_{\rm Img}$, was calculated using the predicted angular separation of the companion at the NIRC2 epoch (2012.55), its apparent $L^{\prime}$ magnitude, and the NIRC2 contrast curve given in Section~\ref{sec:imaging}. We used the interpolated {\tt Sonora} grid to predict the apparent $L^{\prime}$ magnitude of the companion. As the flux of substellar objects is a strong function of their age, we included the age of the system as a free parameter to marginalize over the age uncertainty. A magnitude difference $\Delta L^{\prime}$ was calculated assuming $L^{\prime}=3.8$\,mag for TW PsA derived from empirical spectral type--color relations. $\chi^2_{\rm Img}$ was then calculated as
\begin{equation}
    \chi^2_{\rm Img} = \left(\frac{10^{-0.4\Delta L^{\prime}}}{C\left(\rho\right)}\right)^2,
\end{equation}
where $C(\rho)$ is the value of the 1$\sigma$ contrast curve at the angular separation $\rho$ of the companion in 2012.55. The contrast curve was assumed to be azimuthally symmetric. The value of $C$ interior to the inner working angle and exterior to the outer working angle was fixed to $\infty$, resulting in $\chi^2_{\rm Img}=0$ for companions at these separations.

\subsection{MCMC parameter estimation}
\label{sec:astro_model2}
\begin{deluxetable*}{ccccc}
\tablewidth{0pt}
\tablecaption{MCMC parameters, prior distributions, and fit intervals.\label{tbl:fit2_params}}
\tablehead{\colhead{Parameter} & \colhead{Symbol} & \colhead{Unit} & \colhead{Prior} & \colhead{Prior interval}}
\startdata
Semi-major axis & $a$ & arc sec & Uniform ($\log a$) & $[0.001, 10]$\\
Inclination & $i$ & rad & Uniform ($\cos i$) & $[0, \pi]$\\
Eccentricity & $e$ & \nodata & Uniform & $[0, 1)$\\
Argument of periapse & $\omega$ & rad & Uniform & $[0, 2\pi)$\\
Longitude of ascending node & $\Omega$ & rad & Uniform & $[0, 2\pi)$\\
Mean anomaly at epoch & $\tau$ & \nodata & Uniform & $[0, 1)$\\
Companion mass & $M_2$ & $M_{\odot}$ & Uniform ($\log M_2$) & $[10^{-4}, M_1]$\\
Systemic velocity & $\gamma$ & km\,s$^{-1}$ & Uniform & $[-100, 100]$\\
RV jitter & $\sigma_{\rm jitter}$ & km\,s$^{-1}$ & $\log \sigma_{\rm jitter} \sim \mathcal{N}(-1.6, 0.3^2)$ & $(0, 1]$\\
Age & $t_{\rm age}$ & Myr & $\mathcal{N}(440, 40^2)$ & [100, 780]\\
R.A. offset at 1991.25 & $\Delta\alpha^\star$ & mas & Uniform & $[-50, 50]$\\
Dec. offset at 1991.25 & $\Delta\delta$ & mas & Uniform & $[-50, 50]$\\
Parallax & $\pi$ & mas & Uniform & $(0, 1000]$\\
System R.A. proper motion\tablenotemark{a}& $\mu_{\alpha^{\star}}$ & mas\,yr$^{-1}$ & Uniform & $[281, 381]$\\
System Dec. proper motion\tablenotemark{a}& $\mu_\delta$ & mas\,yr$^{-1}$ & Uniform & $[-209, -109]$\\
\enddata
\tablenotetext{a}{Measured in the {\it Hipparcos} reference frame}
\end{deluxetable*}

This revised astrometric model included fifteen free parameters; the eleven outlined in Section~\ref{sec:astro_model}, the parallax $\pi$, the systemic radial velocity $\gamma$, the amplitude of the radial velocity jitter $\sigma_{\rm jitter}$, and the age of the system $t_{\rm age}$. As in Section~\ref{sec:astro_model}, we used {\tt emcee} to sample the posterior distribution of these fifteen parameters. Table~\ref{tbl:fit2_params} lists the parameters, the assumed prior distribution, and the interval over which the posterior distribution was estimated. Although the prior on the RV jitter parameter was a normal distribution in $\log \sigma_{\rm jitter}$ based on the estimate for the RV jitter given in Section~\ref{sec:rv} ($\log \sigma_{\rm jitter}=-1.6\pm0.3$ [km\,s$^{-1}$]; \citealp{Hillenbrand:2015vp}), we fit for $\sigma_{\rm jitter}$, enforcing a log-normal prior on the parameter. The only difference in choice of priors between this and the previous fit was for the secondary mass. In Section~\ref{sec:astro_model} we used a uniform prior for $M_2$ where $dp/dM_2 \propto1$. Given that the distribution of masses for the wide-orbit giant planet population is more consistent with a power law distribution (e.g., \citealp{Nielsen:2019td}), we instead adopt a uniform prior in the logarithm of the companion mass, $dp/d \log M_2\propto1$ to better match the observed distribution. The prior on the age of the system was a normal distribution using the age estimate of $440\pm40$\,Myr from \citet{Mamajek:2012ga}.

\begin{figure*}
\includegraphics[width=1.0\textwidth]{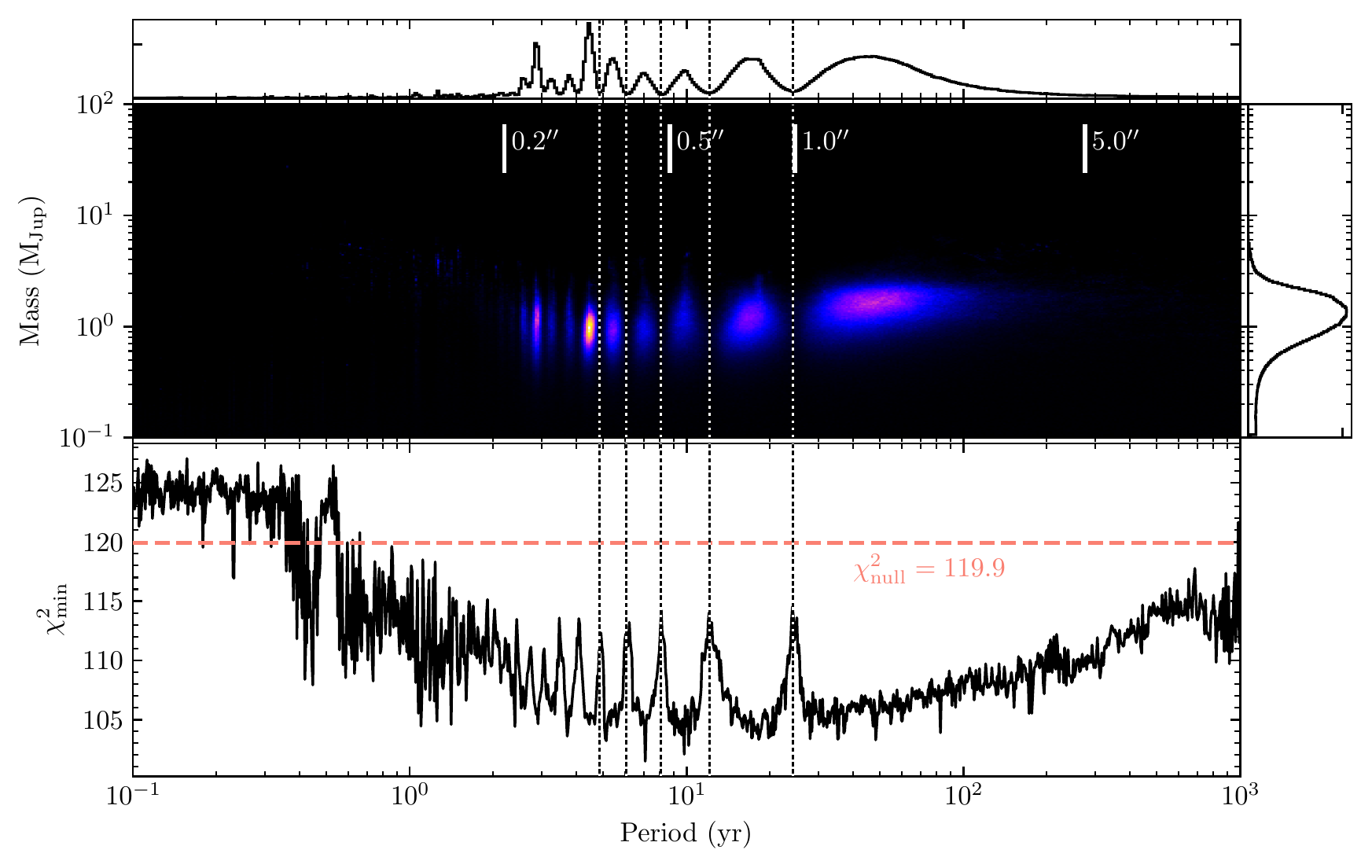}
\caption{As Figure~\ref{fig:instant_prediction} but for the model described in Section~\ref{sec:astro_model2} that incorporates the {\it Hipparcos} IAD, the radial velocity record, and the Keck/NIRC2 contrast curves. The plausible range of companion properties has been significantly reduced relative to the initial fit shown in Figure~\ref{fig:instant_prediction}; the radial velocity record has excluded short-period stellar companions, while the NIRC2 imaging data has excluded substellar companions down to $\sim$3\,M$_{\rm Jup}$ beyond an arcsecond. The fit favors} a long-period ($\gtrsim$3\,yr) Jovian-mass ($M_2=1.2_{-0.6}^{+0.7}$\,M$_{\rm Jup}$) companion.
\label{fig:accl_fit}
\end{figure*}
We initialized 1024 walkers throughout parameter space at each of 20 temperatures, yielding a total of 20480 chains. Each chain was advanced for $10^6$ steps, and the positions were saved every hundredth step. The first half of each chain was discarded as a ``burn-in''. At each step, the log likelihood $\ln \mathcal{L}$ was calculated as 
\begin{equation}
\label{eqn:like2}
\begin{split}
    \ln \mathcal{L} =& -\frac{1}{2}\left(\chi^2_{\rm H} + R_{\rm G}^\top {\mathbf G}_{\rm G}^{-1}R_{\rm G} + \chi^2_{\rm RV} + \chi^2_{\rm Img}\right)\\
    & -\sum\ln\sqrt{2\pi\left(\sigma_v^2 + \sigma_{\rm jitter}^2\right)}.
    \end{split}
\end{equation}
We measured auto-correlation lengths close to unity for each parameter in the decimated chains, suggesting that each saved sample was independent from the previous sample. The median and $1\sigma$ range for each parameter did not significantly change in the remaining half of the chains. The covariance between companion period and mass after marginalizing over the remaining thirteen parameters, and corresponding 1-d marginalized distributions, are shown in Figure~\ref{fig:accl_fit}. Relative to the fit that only includes the absolute astrometry for TW PsA (Figure~\ref{fig:instant_prediction}), the allowed range of parameter space has been significantly constrained. Massive ($\gtrsim$5\,M$_{\rm Jup}$) companions are excluded at short periods by the radial velocity record, and at long periods by the deep NIRC2 imaging data. Furthermore, lower-mass companions with periods shorter than $\sim$2\,yrs are excluded by the radial velocity record. We find a tight constraint on the companion mass $M_2=1.2_{-0.6}^{+0.7}$\,M$_{\rm Jup}$, with a 1, 2, and 3$\sigma$ lower limits of 0.9, 0.2, and 0.1\,M$_{\rm Jup}$ and upper limits of 1.5, 2.4, and 4.5\,M$_{\rm Jup}$. The period is less well constrained ($P=25_{-21}^{+52}$\,yr), with a multi-modal posterior distribution caused by aliasing of the orbit with the 24.25-year baseline between the {\it Hipparcos} and {\it Gaia} missions.

\subsection{Null hypothesis}
We repeated the same exercise described in Section~\ref{sec:null1} where the companion mass is fixed to zero, and the only parameters that are fit are the five astrometric parameters describing the position and motion of the system barycenter ($\Delta\alpha^{\star}$, $\Delta\delta$, $\pi$, $\mu_{\alpha^{\star}}$, $\mu_{\delta}$) and the two parameters describing its radial velocity ($\gamma$, $\sigma_{\rm jitter}$). We used the same MCMC framework described in the previous subsection. We advanced 1024 walkers at each of 20 temperatures for $10^6$ steps, using the likelihood given in Equation~\ref{eqn:like2} and the relevant priors given in Table~\ref{tbl:fit2_params}. The best fit model had $\chi_{\rm null}^2=119.9$, compared with $\chi^2=101.5$ for the model including a massive companion. We calculated the BIC for the model including a massive companion (${\rm BIC} = 171.6$, $k=15$, $n=107$) and the model without (${\rm BIC}_{\rm null} = 152.6$, $k=7$, $n=107$). The $\Delta{\rm BIC}$ of 19.0 corresponds to very strong evidence against the companion hypothesis using the categorization from \citet{Kass:1995eh}, despite the significant reduction in $\chi^2$. The large number of measurements included in the fit leads to a very strong penalty term for each additional parameter used in the model, several of which are effectively nuisance parameters that are marginalized over.

The $\Delta$BIC suggests that the current astrometric measurements are not sufficiently constraining to justify the use of the model described in this section that incorporates a massive orbiting companion over one that assumes linear motion of the star in the sky plane. While there was evidence against the null hypothesis when using this test for the model described in Section~\ref{sec:null1}, the addition of the 87 measurements from the {\it Hipparcos} IAD significantly increased the magnitude of the penalty term in the BIC from 2.1 to 4.7 per additional parameter. Consequently, a significantly smaller $\chi^2$ from the massive companion model in this section is required for there to be significant evidence against the null hypothesis here. A more precise measurement of the astrometric acceleration with future {\it Gaia} data releases will be required in order to reject the null hypothesis at a significant level using the BIC.

\section{Future Prospects for Direct Detection}
The range of orbital periods that satisfy the joint constraints described in the previous section correspond to angular separations ($0\farcs2$--$5\arcsec$) at which the planet could be resolved with current and/or future ground and space-based instrumentation. Two such examples are the Near-Infrared Camera (NIRCam; \citealp{Horner:2004fz}) on the {\it James Webb Space Telescope} ({\it JWST}), and the visible light coronagraphic instrument (CGI; \citealp{Noecker:2016hp}) on the {\it Wide Field Infrared Survey Telescope} ({\it WFIRST}). These two instruments are highly complementary. NIRCam is sensitive to thermal emission from the planet, whilst CGI is sensitive to visible light from the host star reflected off the top of the planet's atmosphere. The youth and proximity of TW PsA, and the evidence presented here suggestive of an orbiting giant planet at a relatively wide angular separation, make it a choice target for these two missions.

\begin{figure}
\includegraphics[width=1.0\columnwidth]{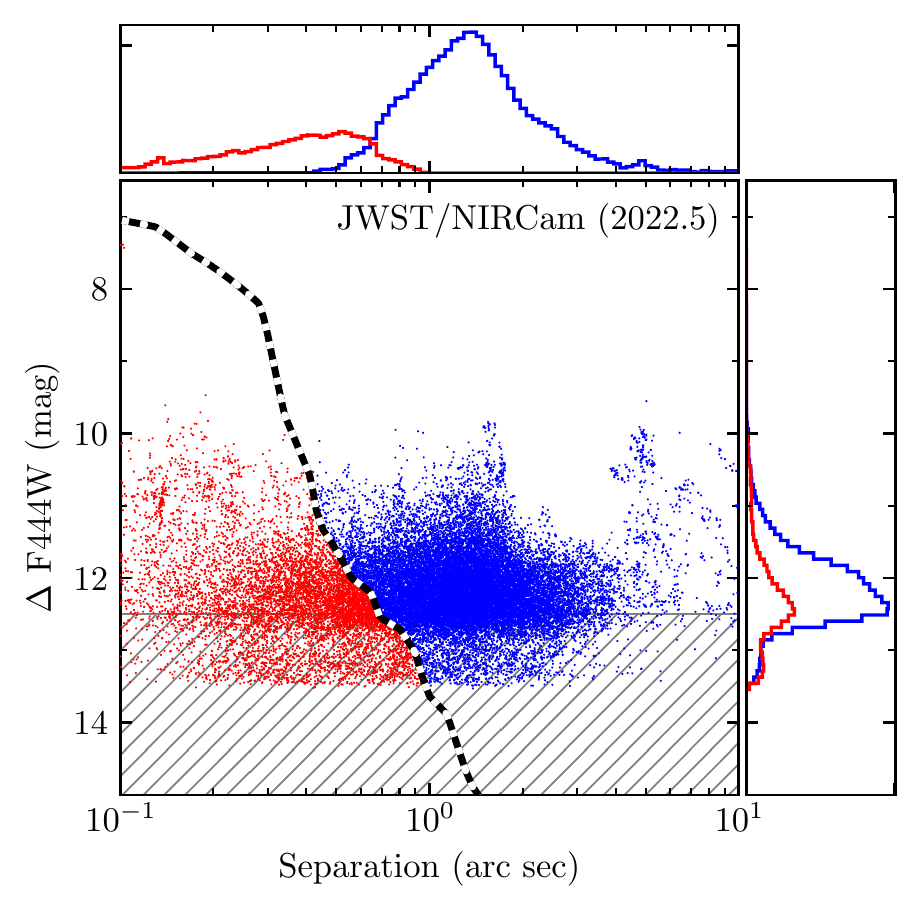}
\caption{Predicted contrast in NIRCam's F444W bandpass as a function of angular separation in 2022.5 for draws from the posterior distributions described in Section~\ref{sec:astro_model2} that lie within the bounds of the {\tt Sonora} grid. Companions that lie above the predicted NIRCam contrast curve (dashed) are plotted in blue, whereas those that are not detectable are plotted in red. The hatched region indicates contrasts at which some planets are beyond the range of the {\tt Sonora} evolutionary and/or atmospheric grids. Marginalized distributions for the detectable and non-detectable samples are also plotted.\label{fig:jw_444w}}
\end{figure}
We used the {\tt Sonora} model grid described previously to predict the flux of the planet in the F444W bandpass, one of the standard filters that will be used in conjunction with coronagraphic imaging with {\it JWST}. We computed the projected separation at 2022.5 for each sample within the MCMC chains from Section~\ref{sec:astro_model2}, and used the corresponding companion mass and system age to predict the flux in the F444W bandpass. We assumed an apparent F444W magnitude for TW PsA of 3.8 based on color transformations computed for main sequence stars. The magnitude difference between star and planet are plotted as a function of angular separation at 2022.5 in Figure~\ref{fig:jw_444w}. We find that 24\,\% of the draws from the posterior distributions lie above the predicted contrast of NIRCam \citep{Beichman:2010ia}. A significant fraction (68\,\%) of the MCMC samples were for masses and ages that fell outside of the {\tt Sonora} evolutionary and/or atmospheric grids, typically due to the predicted temperature for the planet being beyond the range of the atmospheric grid. We cannot assess their detectability with the current model grid, but a small subset at the widest angular separations will lie above the predicted contrast curve unless a significant decrease in flux at 4.5\,$\micron$ occurs at temperatures below 200\,K.

The proximity of TW PsA (7.6\,pc) and the small angular separation make this target particularly favourable for the direct detection of the reflected light of the host star from the top of the planet's atmosphere. We used a grid of reflectance spectra for giant planets \citep{Batalha:2018cz} to predict the contrast between the star and planet at 575 and 825\,nm, two of the filters proposed for the CGI instrument. Within this model grid the reflectivity is expressed as the geometric albedo $p$ scaled by the phase function $\Phi$ at a given orbital phase angle $\beta$, and depends on the planet metallicity $[M/H]$, separation $r$, cloud sedimentation parameter $f_{\rm sed}$ \citep{Ackerman:2001gk}, and the wavelength of the observations. We calculated orbital separations and phase angles at 2027.0 for each MCMC sample, and estimated radii using an empirical mass-radius relationship\footnote{\url{https://plandb.sioslab.com/docs/html/}} \citep{Chen:2017fg}. As the metallicity and cloud properties of the planet are not known, we calculate star to planet contrasts at each grid point within the ($[M/H]$, $f_{\rm sed}$) plane to evaluate detectability over a representative range of planet properties. The contrast and angular separation at 2027.0 was then compared to predicted CGI sensitivity curves\footnote{\url{https://github.com/nasavbailey/DI-flux-ratio-plot}} at 575 and 825\,nm assuming 100\,hr of on-source integration \citep{Nemati:2017kv}. We note that the mass-radius relationship and the reflectence model are not coupled; the radius of a solar composition planet of a given mass is assumed to be the same as one with an enhanced abundance. In reality, these higher metallicity planets would have smaller radii, and our model overestimates their detectibility.

\begin{figure}
\includegraphics[width=1.0\columnwidth]{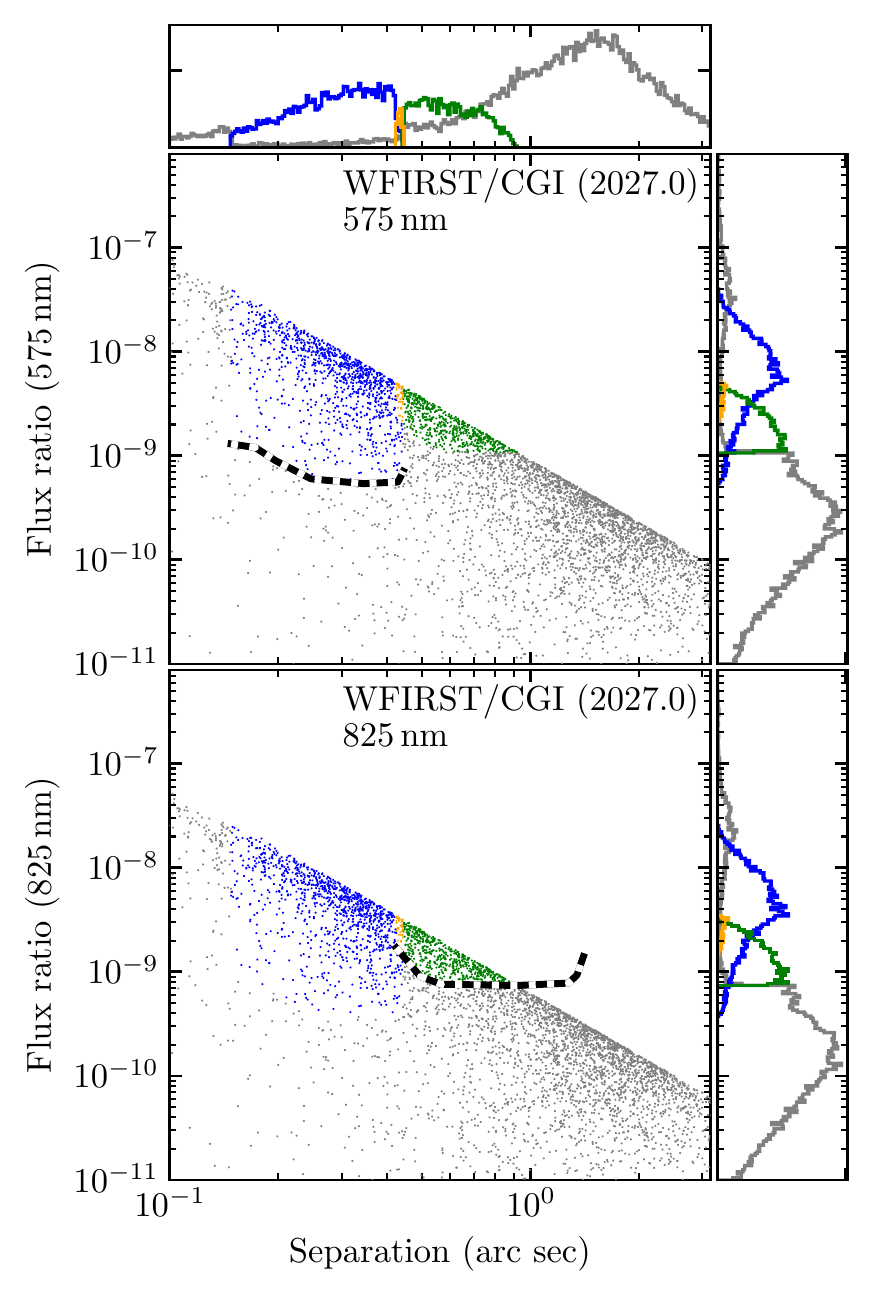}
\caption{Reflected light contrast at 575\,nm (top panel) and 825\,nm (bottom panel) as a function of angular separation in 2027 for draws from the posterior distributions described in Section~\ref{sec:astro_model2}. Planets detectable in 100\,hr with the 575\,nm filter (blue), the 825\,nm filter (green), or with both filters (yellow) are highlighted. Predicted sensitivity curves for imaging observations with {\it WFIRST}/CGI are overplotted (dashed lines), limited by angular size of the control region of the deformable mirror in the focal plane. Corresponding marginalized distributions are also shown.\label{fig:cgi1}}
\end{figure}
\begin{figure}
\includegraphics[width=1.0\columnwidth]{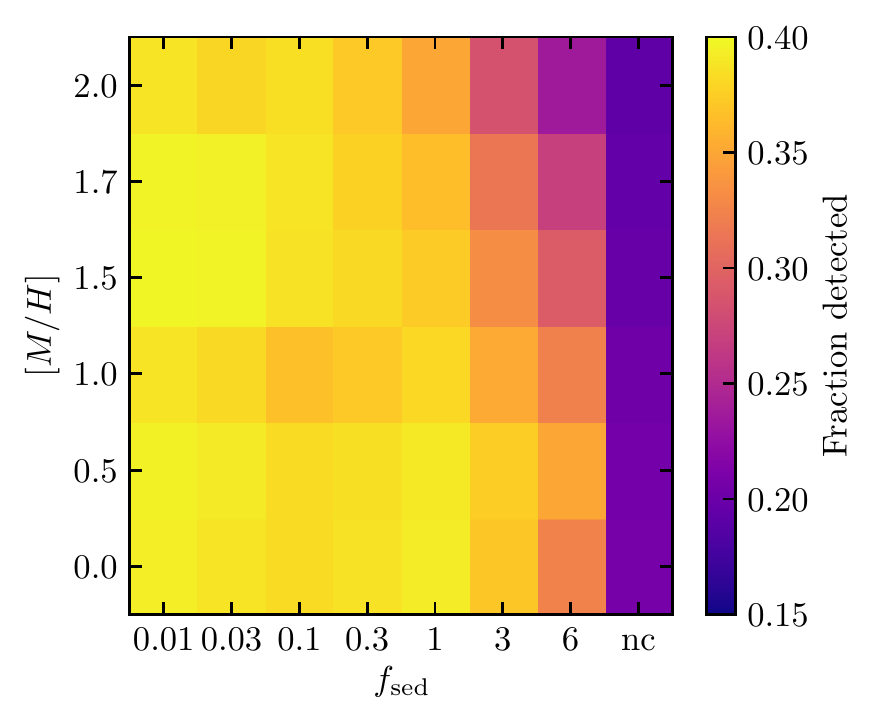}
\caption{Fraction of planets from the MCMC posterior distributions that are detectable with {\it WFIRST}/CGI as a function of the cloud sedimentation parameter $f_{\rm sed}$ and metallicity $[M/H]$. Detection limits calculated assuming 100\,hr exposure time per filter. \label{fig:cgi2}}
\end{figure}
The predicted contrast as a function of angular separation in 2027.0 is shown for the 575\,nm and 825\,nm filters in Figure~\ref{fig:cgi1} assuming $[M/H] = 0$ and $f_{\rm sed} = 6$ (thin clouds), a median scenario in terms of planet detectability. The strong dependence of the planetary albedo on the cloud sedimentation parameter, and to a lesser extent the metallicity, leads to a large variation in the probability of detection as a function of these parameters (Figure~\ref{fig:cgi2}). With efficient sedimentation (large values of $f_{\rm sed}$) the clouds are vertically thin, resulting in a relatively low albedo. As $f_{\rm sed}$ is decreased the clouds become thicker, resulting in a higher albedo. Low values for this sedimentation parameter have been required to explain observations of the atmospheres of transiting hot Jupiters (e.g., \citealp{Demory:2013fr}), whereas more widely separated gas giants like Jupiter might be limited to $f_{\rm sed} \ge 3$ \citep{Batalha:2018cz}. Assuming a solar metallicity, going from the cloud-free albedo spectra to one with thin clouds ($f_{\rm sed}= 6$) increases the fraction of planets that are detectable with CGI from 20\,\% to 29\,\% due to the significant increase in the geometric albedo.

\section{Conclusions}
We have presented the first constraints on planetary-mass companions to the nearby (7.6\,pc) young ($\sim$440\,Myr) K4Ve star TW Piscis Austrini, a wide common proper motion companion to Fomalhaut, that combine absolute astrometry, direct imaging, and radial velocities. Previous studies had identified this star as exhibiting an astrometric acceleration between the {\it Hipparcos} and {\it Gaia} missions \citep{Kervella:2019bw}. While the significance of the acceleration is $\le4\sigma$ in each pair of derived proper motions, the goodness of fit of the measurements to an astrometric model describing the motion of the star across the sky is significantly worse without accounting for the reflex motion induced by a massive orbiting companion. We combined these absolute astrometric measurements with Keck/NIRC2 $L^{\prime}$ coronagraphic imaging and an 11-year radial velocity record \citep{TalOr:2019gm,Butler:2017km} that exclude massive ($\gtrsim5$\,M$_{\rm Jup}$ $3\sigma$ limit) companions over all plausible orbital periods. The combination of this upper limit with a lower limit derived from the significant astrometric acceleration leads to a tight constraint on the mass of a companion consistent with the magnitude and direction of the acceleration of $M_2=1.2_{-0.6}^{+0.7}$\,M$_{\rm Jup}$. The orbital period is less constrained ($P=25_{-21}^{+52}$\,yr), and is aliased with the 24.25-year baseline between the {\it Hipparcos} and {\it Gaia} missions. Continued radial velocity monitoring and deep imaging observations will further constrain the properties of this putative companion, while the upcoming {\it Gaia} data releases will more precisely measure the acceleration of the host star.

Searches for exoplanets via absolute astrometry using ground-based instrumentation have yielded either null results or candidates that were later found to be spurious (e.g., 61 Cygni, \citealp{Strand:1943dk,Wittenmyer:2006jx}; Van Biesbroeck's star, \citealp{Pravdo:2009ko,Lazorenko:2011hc}). Such surveys are limited by the astrometric precision of the individual measurements, typically $>0.1$\,mas \citep{Lazorenko:2009ju}. There have also been numerous attempts to measure the astrometric signal of known exoplanets with the Fine Guidance Sensors on the {\it Hubble Space Telescope} (e.g., \citealp{Benedict:2017gc}); however, several of these measurements were later found to be inconsistent with analyses incorporating additional radial velocity data (e.g., \citealp{Rivera:2010gi, Mawet:2019cd}).  The exquisite precision of {\it Gaia}, an order of magnitude better than current ground-based instrumentation, is set to transform this field of study with thousands of exoplanets detected via astrometry alone \citep{Perryman:2014jr}, the first since the discovery of Neptune in the mid-19th century.

\acknowledgments
The authors wish to thank Dmitry Savransky and Vanessa Bailey for useful discussions relating to this work. The authors were supported in part by NSF AST-1411868 (R.D.R., E.L.N., B.M.) AST-1518332 (R.D.R., T.M.E., P.K., J.J.W), NASA NNX14AJ80G (R.D.R., E.L.N., B.M.) NNX15AC89G (R.D.R., T.M.E., P.K., J.J.W), NNX15AD95G (R.D.R., T.M.E., P.K., J.J.W), NSSC17K0535 (R.D.R., E.L.N., B.M.), and NNG16PJ24C (R.D.R., B.M., E.L.N.). J.J.W. is supported by the Heising-Simons Foundation 51~Pegasi~b postdoctoral fellowship. This work benefited from NASA's Nexus for Exoplanet System Science (NExSS) research coordination network sponsored by NASA’s Science Mission Directorate. Some of the data presented herein were obtained at the W. M. Keck Observatory, which is operated as a scientific partnership among the California Institute of Technology, the University of California and the National Aeronautics and Space Administration. The Observatory was made possible by the generous financial support of the W. M. Keck Foundation. The authors wish to recognize and acknowledge the very significant cultural role and reverence that the summit of Maunakea has always had within the indigenous Hawaiian community.  We are most fortunate to have the opportunity to conduct observations from this mountain. This work has made use of data from the European Space Agency (ESA) mission {\it Gaia} (\url{https://www.cosmos.esa.int/gaia}), processed by the {\it Gaia} Data Processing and Analysis Consortium (DPAC; \url{https://www.cosmos.esa.int/web/gaia/dpac/consortium}). Funding for the DPAC has been provided by national institutions, in particular the institutions participating in the {\it Gaia} Multilateral Agreement. This research has made use of the SIMBAD database and the VizieR catalog access tool, both operated at the CDS, Strasbourg, France. This research has made use of the Imaging Mission Database, which is operated by the Space Imaging and Optical Systems Lab at Cornell University.

\facility{Keck:II (NIRC2)}

\software{Astropy \citep{TheAstropyCollaboration:2013cd},  
          Matplotlib \citep{Hunter:2007ih},
          orbitize \citep{Blunt:2019vq}, emcee \citep{ForemanMackey:2013io}, radvel \citep{Fulton:2018gq}.}

\bibliographystyle{aasjournal}

\end{document}